# A model for liquid-striped liquid phase separation in liquids of anisotropic polarons

D. Innocenti[1], A. Ricci[1], N. Poccia[1], G. Campi[2], M. Fratini[1], Antonio Bianconi[1*]

[1]*Department of Physics, Sapienza University of Rome, P. Aldo Moro 2, 00185 Rome, Italy*
[2]*Istituto di Cristallografia, CNR, Via Salaria Km 29, 300, 00016 Monterotond,Rome, Italy*

**Abstract:** The phase separation between a striped polaron liquid at the particular density and a high density polaron liquid is described by a modified Van der Waals scheme. The striped polaron liquid represents the pseudo gap matter or Wigner-lke polaron phase at 1/8 doping in cuprate superconductors. The model includes the tendency of pseudo-Jahn Teller polarons to form anisotropic directional bonds at a preferential volume with the formation of different "liquid phases". The model gives the co-existence of the polaron striped-liquids – high density liquid that appears in cuprate superconductors for doping larger than 1/8. We discuss as the strength of anisotropic bonds controls the variation the phase separation scenarios for complex systems with the presence of quantum critical point where the phase separation vanishes.

*Corresponding author Antonio Bianconi*
*Email Antonio.bianconi@roma1.infn.it*

1. **Introduction**

The Van der Waals scheme is the simplest scheme to describe the spinodal phase separation in complex system such as polaron liquids. It has been used to study the phase separation in a polaron liquid by Emin [1,2]. We propose to extend the Van der Walls scheme to describe the complex phase separation in cuprates appearing for doping larger than 1/8 where a polaron striped metallic phase with doping larger than 1/8 co-exists with a high density phase [3-6]. This phase separation has been shown to be driven by the co-existence of pseudo Jahn-Teller polarons (PJTP), associated with the rombic distortion of the $CuO_4$ square plane and its tilting in the vertical o horizontal direction or antinodal directions.[7-10] These PJTP polarons are in the intermediate coupling range between large and small polarons, having an area of about 8 Cu sites. [11] These PJTP coexist with itinerant particles moving in the nodal direction. [12,13]

The PJTP get self organized below a critical temperature T* with the formation of electronic polaron striped phases at critical density close to 1/8 holes per Cu site, giving the so called pseudogap phase. [14-21] The interaction between the PJTP is clearly anisotropic because of the associated rombic distortion therefore they have the tendency to form linear polaron strings. [22, 23] The polaronic electron-phonon interaction depends in these heterostructrures at the atomic limit [24-26] on the lattice misfit strain between the active $CuO_2$ layers and the spacer layers. [27-31] that plays a key role also in diborides [32] and iron pnictides [33].

The phase diagram of cuprate superconductors shows that a PJTP polarons striped phase called also a quasi 1D generalized Wigner polaron crystal is formed at critical values of the charge density (1/8 holes per Cu site and a critical strength of the elastic field due to the misfit strain. The critical density for the formation of a electronic polaronic crystal depends on the effective volume occupied by the polaron carriers. The anisotropic interactions between the polarons produce the stripe phases. In order to give account of the new emerging complex phase diagram, where phase separation is controlled by doping and misfit strain [27-33], we have extended the model of Poole et al. [34] for supercooled water, as proposed by Campi et al. [35], to describe the 3D phase diagram including all different HTcS families.

2. **The model**

We use a modified Van der Waals interaction model, analogously to the one introduced for the phase diagram of supercooled water [34,35]. In fact, supercooled water, a prototype of complex matter, shows a phase separation driven by the tendency of water molecules to form quasi 1D arrays of hydrogen bonds. This tendency gives fluctuating clusters made of a low density liquid (LDL) that coexists with the high density liquid water (HDL). In order to describe this anomalous phase separation in water, the model of Poole et al. [35] implements the standard Van der Waals model by including a characteristic gain in energy "$\gamma$" for the formation of clusters with directional hydrogen bonds at a particular preferential volume "$V_\gamma$". The introduction of this anisotropic interaction provides a phase diagram with the coexistence of a high density liquid (HDL) and a low density liquid (LDL) when $\gamma$ is larger than a threshold value.

Here we have extended the Poole model in order to describe the phase separation in the polaron liquid in cuprates. The free energy of a complex system in which a phase separation is observed, is obtained by adding the term $A_\gamma$, to the Van der Waals free energy $A_{VDW}$ yielding a total free energy:

$$A = A_{VDW} + A_\gamma \qquad (1)$$

where $A_{VdW}$ and $A_\gamma$ are given by:

$$A_{VdW} = -RT\{\ln[(V-b)/\Lambda^3] + 1\} - a^2/V \qquad (2)$$

$$A_\gamma = -RTf \ln\left[\Omega + \exp\left(-\frac{\gamma}{RT}\right)\right] - RT(1-f)(\Omega+1) \qquad (3)$$

Here *a* and *b* are the standard Van der Waals constants. The Van der Waals constant *a* is associated with the isotropic inter-particle attraction and the constant *b* is associated to the volume occupied by fluid particle. These parameter are used to define the critical values of thermodynamic variables $T_c$, $P_c$ and $V_c$ and the standard parameter $\gamma_0 = a/b$ for the Van der Waals model. The value of the parameter $\Omega$ is defined as $\Omega = \exp(-S_\gamma/R)$, where $S_\gamma$ is entropy of formation for a mole of anisotropic bonds. The intermediate phase is characterized by the anisotropic inter-particle interaction $\gamma$. In this approach there are $\Omega \gg 1$ configurations all having $\gamma = 0$ and only a single configuration in which the formation of the anisotropic bonds with energy $\gamma$ is allowed. The anisotropic interaction is most likely to occur when the bulk molar volume is consistent with the preferential volume $V_\gamma$. In fact, each particle has an optimal local volume for the formation of anisotropic bonds to its neighbors. Changing the particle density, when $V \neq V_\gamma$, the anisotropic interactions are only a fraction *f* of the total, since V is no longer consistent with the possibility that all anisotropic interactions are saturated at the optimal volume. The remaining fraction of bonds, 1-*f*, occurs in an unfavorable local volume and therefore, they cannot form the anisotropic bonds of the phase. The term *f* is given by:

$$f = \left[-\left(\frac{V-V_\gamma}{\sigma}\right)^2\right] \qquad (4)$$

where $\sigma$ characterizes the width of the region of volume around $V_\gamma$ over which a significant fraction of anisotropic bonds can be described by Eq. (3).

To describe the phase diagram of xuprates where we observe two phase separations, we consider a Poole model with one anisotropic interaction $\gamma$ giving rise to the intermediate phase with preferential volume $V_\gamma$ that simulate the striped polaron phase in cuprates at doping 1/8.

The inclusion in the model of the optimum volume $V_\gamma$ for the anisotropic bond introduces a new minimum in the free energy when V approaches $V_\gamma$. In Fig. 1 we have plotted the normalized free energy $A/\gamma_0$ as functions of the normalized volume $V/V_c$ at the fixed normalized interaction energy $\gamma/\gamma_c = 1$. We observe in the phase diagram that the new minimum occurs only for temperature lower than the critical value $T/T_c = 0.68$.

We observe that the effect of the $A_\gamma$ term in Eq. (1), related with the strength of the anisotropic interactions, is to "split" the normal Van der Waals spinodal curve by imposing thermodynamic stability in the region of states centered at the reduced density $b/V = 0.5$ where the intermediate phase is stable. As a result, two spinodal lines occur, each terminating at a critical point producing two phase separation regions. As the directional bond energy $\gamma$ decrease respect with the Van der Waals interaction $\gamma_0$, the phase separations generated by the strength of the directional bond decrease and the stabilizing effect of $A_\gamma$ set in only at lower temperature. The critical points merge with the high density spinodal of the main Van der Waals spinodal line that is formed when $\gamma$ go to zero.

In the phase diagrams of panel a), b), c) of Fig. 2 we change the $\gamma$ value in order to show the effects of the strength of the anisotropic interaction on the thermodynamic behavior of the system.

In panel a), we have used the value $\gamma/\gamma_0 = 3.0$, in fact for a value of $\gamma/\gamma_0 > \gamma_c$ we observe that appear a second phase separation that becomes more defined when we continue to increase $\gamma$ interaction.

When $\gamma$ decreasing to the critical value ($\gamma=\gamma_c$) $\gamma/\gamma_0 = 1.48$, the two spinodal lines, (1+2) and (2+3), are going to overlap in a point (panel b), and then, when $\gamma$ becomes lower than the threshold $\gamma_0$, a phase separation merges with the other in a greater region of phase coexistence and the intermediate phase (2) becomes an isolated pocket of stability completely enclosed within a spinodal line as show in the panel c). If we continue to decrease the strength of the anisotropic interaction we obtain a more isolated pocket with a lower critical temperature until when the anisotropic interaction $\gamma/\gamma_0$ go to zero we reach a quantum critical point where we obtain the standard Van der Waals phase diagram.

## 3. Interaction energy phase diagram

Here present the phase diagram at constant low temperature as a function of the directional interaction energy and the reduced density. We have used the previous phase diagrams to get the new one, in fact we calculate the value of each interaction energy normalized $\gamma/\gamma_c$ at a certain reduced density b/V with the temperature fixed at a fixed value.

As we have normalized all the thermodynamic variables and energies respect with their critical values we obtain a universal phase diagram.

We obtain that three different phases occur, separated by three regions of phase coexistence. As the temperature increases, the phase separations generated by the strength of the directional bond decrease and the different phases are no longer well separated.

In the phase diagram of Fig. 3 we use value $T/T_c = 0.4$ and we observe the occurrence of three different phases, well separated by three regions of phase coexistence.

## 4.     Conclusions

In conclusion we have presented a model for an electronic complex system that simulate the phase diagram of the polaron liquid in cuprates with coexistence of different electronic phases at critical densities and coexistence of different liquids described by the modified Van der Waals model.

We discuss the critical values of the anisotropic interactions for the spinodal lines and we find that this model is able to describe a generic complex system with a variable anisotropic energy interaction.

Finally we have shown a phase diagram for complex system where critical temperature depends on the density and the energy of the anisotropic interaction.

This approach might be useful to study complex phased separations like that in high $T_c$ superconductors

**Aknowledgements:** This work was supported by European project 517039 "Controlling Mesoscopic phase separation" (COMEPHS) (2005).

# References

1. D. Emin, Phys. Rev. B, 49, 9157 (1994).
2. D. Emin, Phys. Rev. Lett. 72, 1052 (1994).
3. F. V. Kusmartsev, D. Di Castro, G. Bianconi and A. Bianconi, *Transformation of strings into an inhomogeneous phase of stripes and itinerant carriers* Physics Letters A 275, 118 (2000).
4. F. V. Kusmartsev and Mikko Saarela *Two-component physics of cuprates and superconductor–insulator transitions* Supercond. Sci. Technol. 22, 014008 (2009)
5. K. I. Kugel, A. L. Rakhmanov, A. O. Sboychakov, Nicola Poccia and Antonio Bianconi *A model for the phase separation controlled by doping and the internal chemical pressure in different cuprate superconductors*. Phys. Rev. B 78, 165124 (2008)
6. K I Kugel, A L Rakhmanov, A O Sboychakov, F V Kusmartsev, Nicola Poccia and Antonio Bianconi *A two-band model for the phase separation induced by the chemical mismatch pressure in different cuprate superconductors* Supercond. Sci. Technol. 22, 014007 (200
7. A. Bianconi, A.-M. Flank, P. Lagarde, C. Li, I. Pettiti, M. Pompa, and D. Udron *On the Cu 3d holes in high Tc superconductors. Is a Cu $3d_{z^2-r^2}$ bipolaron the superconducting pair?* in "Electronic Structure and Mechanisms of High Temperature Superconductivity", J. Askenazi, and G. Vezzoli editors, Plenum, New York (1992) (Proc. of the Miami Workshop, Miami 3-9 January 1991.
8. A. Bianconi N. L Saini, A. Lanzara, M. Missori, T. Rossetti, H. Oyanagi, H. Yamaguchi, K. Oka, T.Ito *Phys. Rev. Lett* 76, 3412-3415 (1996)
9. K. A. Müller *Essential Heterogeneities in Hole-Doped Cuprate Superconductors* in "Superconductivity in Complex Systems" (Springer, Berlin) Book Series Structure & Bonding 114, 1-11 (2005) DOI 10.1007/b101015
10. K. A. Müller *From Phase Separation to Stripes* in "Stripes and Related Phenomena" edited by A. Bianconi and N. L. Saini, (Kluwer – Plenum, New York, 2002) Book Series: Selected Topics in Superconductivity vol. 8, 1-8 (2002) DOI 10.1007/b119246
11. A. Bianconi, M. Missori, H.Oyanagi, and H. Yamaguchi D. H. Ha, Y. Nishiara and S. Della Longa *The measurement of the polaron size in the metallic phase of cuprate superconductors* Europhysics Letters 31, 411 (1995)
12. K. A. Müller, G.-M. Zhao, K. Conder and H. Keller Journal of Physics: Condensed Matter 10, L291 (1998)
13. K. M. Shen, F. Ronning, D.H. Lu, F. Baumberger, N.J.C. Ingle, W.S. Lee, W. Meevasana, Y. Kohsaka, M. Azuma, M. Takano, H. Takagi, Z.X. Shen, Science 307, 901 (2005).
14. M. Missori, A Bianconi, N. L. Saini and H. Oyanagi *High Critical Temperature by Resonant Quantum Confinement: Evidence for Polarons Ordering at T*~1.5Tc in Bi-2212 and La-214 by EXAFS* Il Nuovo Cimento D 16, 1815 (1994)
15. A. Bianconi and M. Missori *The Coupling of a Wigner Charge Density Wave with Fermi Liquid from the Instability of a Wigner Polaron Crystal: A Possible Pairing Mechanism in High Tc Superconductors* in "Phase Separation in Cuprate Superconductors", E. Sigmund and K. A. Müller, editors Springer Verlag, Berlin-Heidelberg,. 272-289 (1994). Proc. of the Workshop held at Cottbus, Germany Sept 4-10, 1993
16. G. Campi, D. Di Castro, G. Bianconi, S. Agrestini, N. L. Saini, H. Oyanagy, and A. Bianconi *Photo-Induced phase transition to a striped polaron crystal in cuprates* Phase Transitions, 75, 927 (2002)

17. A. Bianconi, D. Di Castro, G. Bianconi, A. Pifferi, N. L. Saini, F. C. Chou, D. C. Johnston, and M. Colapietro, *Coexistence of stripes and superconductivity: Tc amplification in a superlattice of superconducting stripes*, Physica C 341, 1719 (2000).
18. D. Di Castro, M. Colapietro and G. Bianconi, *Metallic stripes in oxygen doped La2CuO4* International Journal of Modern Physics B 14, 3438 (2000). DOI:10.1142/S0217979200003927
19. A. Bianconi *The Instability Close to the 2D Generalized Wigner Polaron Crystal Density: A possible Pairing Mechanism Indicated by a Key Experiment* Physica C 235-240, 269 (1994)
20. A. Bianconi *On the Fermi Liquid Coupled with a Generalized Wigner Polaronic CDW Giving High Tc Superconductivity* Sol. State Commun. 91, 1 (1994)
21. A. Bianconi, M. Missori *The instability of a 2D Electron Gas Near the Critical Density for a Wigner Polaron Crystal Giving the Quantum State of Cuprate Superconductors* Sol. State Commun. 91, 287 (1994).
22. F. V. Kusmartsev Phys. Rev. Lett. 84, 530 (2000); ibidem 84, 5026 (2000).
23. F. V. Kusmartsev, Europhys. Lett. 54, 786 (2001).
24. A. Bianconi "*Process of increasing the critical temperature Tc of a bulk superconductor by making metal heterostructures at the atomic limit*" United State Patent No.:US6, 265, 019 B1, (2001).
25. Y. Tokura and T. Arima *"New Classification Method for Layered Copper Oxide Compounds and Its Application to Design of New High $T_c$ Superconductors"* Jpn. J. Appl. Phys. 29 2388 (1990) DOI: 10.1143/JJAP.29.2388
26. A. Bianconi *On the possibility of new high Tc superconductors by producing metal heterostructures as in the cuprate perovskites* Solid state communications 89, 933 (1994).
27. M. Fratini, N. Poccia, and A. Bianconi *Journal of Physics Conference Series* 108, 012036 (2008).
28. A. Bianconi, N. L. Saini, S. Agrestini, D. Di Castro, and G. Bianconi Int. Jour. of Modern Physics B 14, 3342 (2000).
29. S. Agrestini, N L Saini, G Bianconi, and A Bianconi *The strain of CuO2 lattice: the second variable for the phase diagram of cuprate perovskites* Journal of Physics A: Mathematical and General 36, 9133 (2003).
30. N. Poccia and M. Fratini *"The Misfit Strain Critical Point in the 3D Phase Diagrams of Cuprates"* Journal of Superconductivity and Novel Magnetism, 22, 1557 (2009). DOI 10.1007/s10948-008-0435-8.
31. A. Bianconi, G. Bianconi, S. Caprara, D. Di Castro, H Oyanagi, and N. L. Saini, J. Phys.: Condens. Matter, 12 10655 (2000).
32. S. Agrestini et al. *"High $T_c$ superconductivity in a critical range of micro-strain and charge density in diborides"* J. Phys.: Condens. Matter 13, 11689 (2001)
33. Rocchina Caivano, et. al. *Feshbach resonance and mesoscopic phase separation near a quantum critical point in multiband FeAs-based superconductors* Supercond. Sci. Technol. 22, 014004 (2009).
34. P. H. Poole, F. Sciortino, T. Grande, H. E. Stanley, and C. A. Angell, Phys. Rev. Lett. 73, 1632 (1994) and references therein.
35. G. Campi and A. Bianconi, in Symmetry and Heterogeneity in High Temperature Superconductors (Springer, Dordrecht, The Netherlands) *NATO Science Series II Mathematics, Physics and Chemistry* Vol. 214, 147 (2006).

**Figure captions**

**Figure 1.** The normalized free energy as a function of reduced volume $V/V_c$ at different temperatures. In this way the novel minimum of the free energy occurs at a reduced density different respect the minimum of the VdW model. This new minimum, indicates the occurrence of the intermediate phase, become deeper when the temperature decreases, as shown in the plots for $T/Tc$ = 0.40, 0.44, 0.48, 0.56, 0.60, 0.68 and 0.76.

**Figure 2.** The phase diagrams obtained by computing the spinodal lines from Eq. (1). We can observe the occurrence of two phase separations indicated by the two spinodal lines in panel a). In the panel b) and c) we show the effect of the decreasing strength of the directional bond.

**Figure 3.** The normalized energy of the directional interaction $\gamma/\gamma_c$ as a function of reduced density $b/V$ at $T/T_c$=0.4. We identify three different phases, well separated by three regions of phase coexistence with a critical minimum value of the directional interaction $\gamma/\gamma_c,$ below this minimum value the a pure phase (2) assigned to the PJT striped polaron phase, or pseudo gap phase, is not observed.

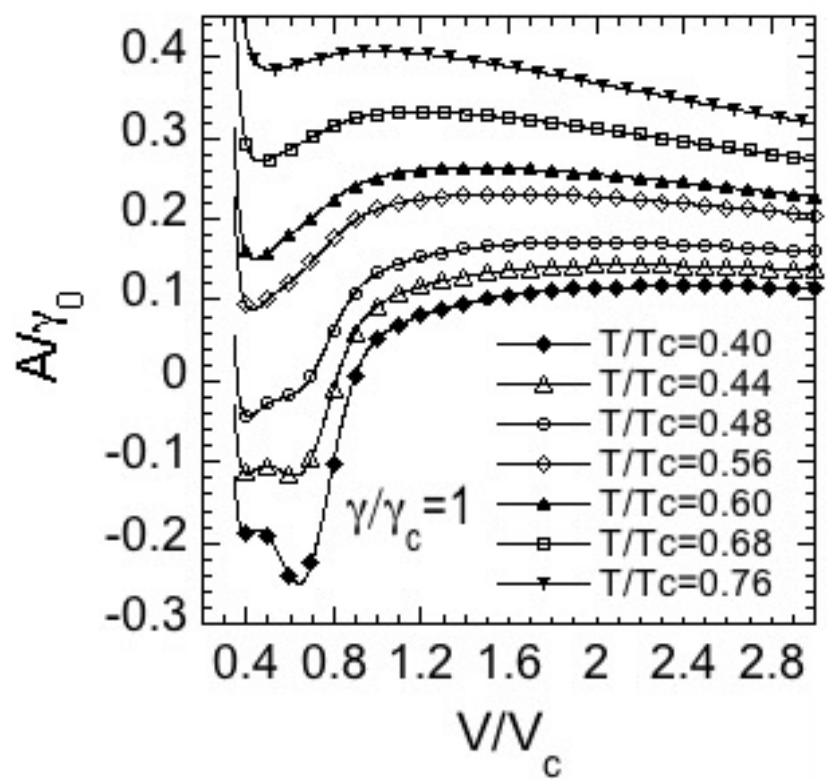

**Figure 1.**

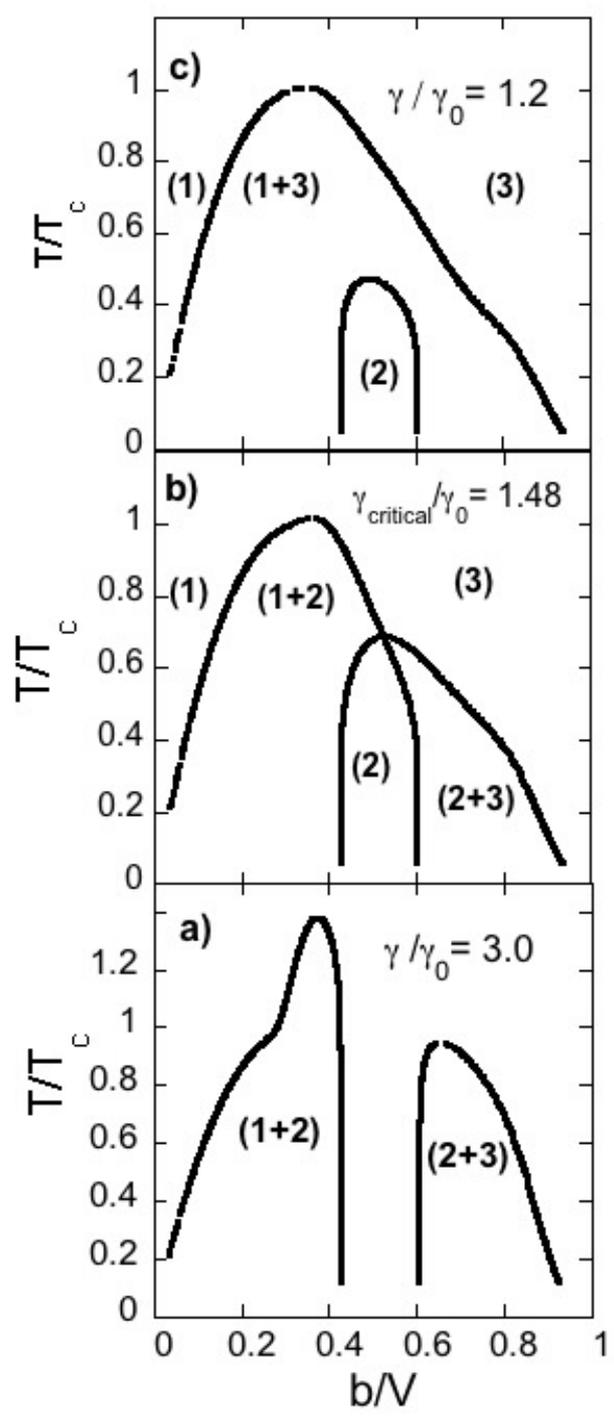

**Figure 2**

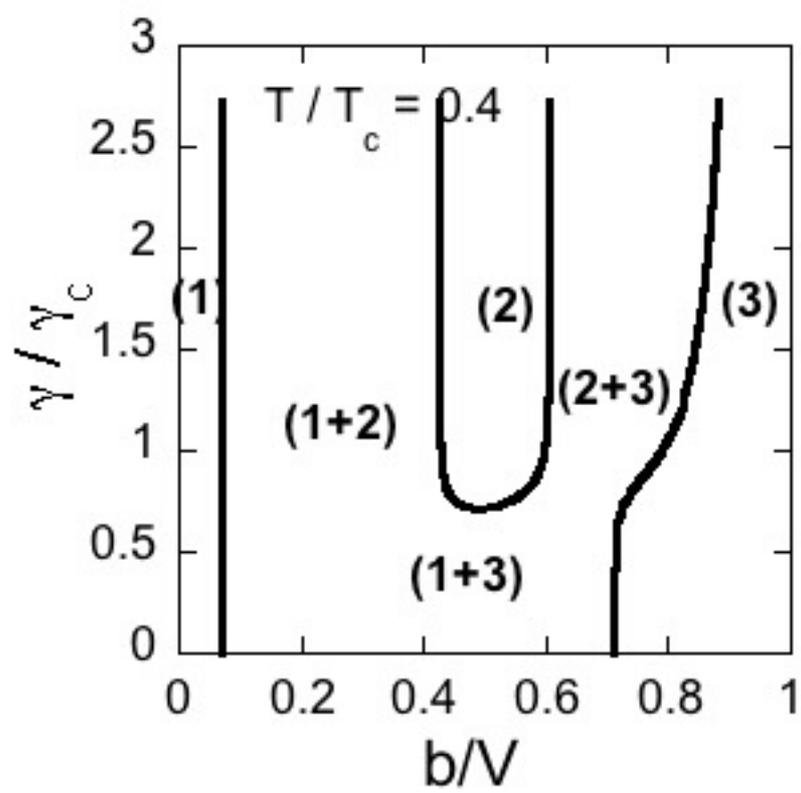

**Figure 3.**